\documentclass[twocolumn,aps,prl,showpacs,preprintnumbers,amsmath,amssymb]{revtex4}
\usepackage{graphicx}
\usepackage{bm}
\usepackage{gensymb}
\usepackage{color}
\begin{document}

\title{ Orbital ordering of Ir-t$_{2g}$ states in the double perovskite Sr$_2$CeIrO$_6$}

\author{Sudipta Kanungo$^1$, Kailash Mogare$^1$, Binghai Yan$^{1,2}$, Claudia Felser$^1$, and Martin Jansen$^{1,3,*}$}
\affiliation{ $^1$Max-Planck-Institut fur Chemische Physik fester Stoffe, 01187 Dresden, Germany\\
$^2$ Max-Planck-Institut fur Physik komplexer Systeme, 01187, Dresden, Germany\\
$^3$ Max-Planck-Institut fur Festkorperforschung, 70569 Stuttgart, Germany}
\date{\today}                                         

\begin{abstract}
The electronic and magnetic properties of monoclinic double perovskite Sr$_2$CeIrO$_6$ were examined based on both experiments and first-principles 
density functional theory calculations. From the calculations we conclude that low-spin-state Ir$^{4+}$ (5$\textit{d}^5$, S=$\frac{1}{2}$) shows  
t$_{2g}$ band derived anti-ferro type orbital ordering implying alternating occupations of $\textit{d}_{yz}$ and $\textit{d}_{xz}$ orbitals at the two symmetrically independent Ir sites. The experimentally determined Jahn-Teller type distorted monoclinic structure is consistent with the proposed orbital ordering picture. Surprisingly, the Ir-5$\textit{d}$ orbital magnetic moment was found to be $\approx$ 1.3 times larger than the spin magnetic moment. 
The experimentally observed AFM-insulating  states are consistent with the calculations. Both electron-electron correlation and spin-orbit coupling (SOC) are required to drive the experimentally observed AFM-insulating ground state. This single active site double perovskite provides a rare platform with a prototype geometrically frustrated fcc lattice where among the different degrees of freedom (i.e spin, orbital, and lattice), spin-orbit interaction and Coulomb correlation energy scales compete and interact with each other.

\end{abstract}

\maketitle{}
Transition metal oxides (TMOs) with orbitally degenerate $\textit{d}$ electrons exhibit rich electronic and magnetic properties arising from the 
interplay of spin, orbital, charge, and lattice degrees of freedom. When the number of $\textit{d}$ electrons (or holes) in a transition metal 
is smaller than the degeneracy of $\textit{d}$ orbitals, the $\textit{d}$ electrons have the freedom to occupy any of them. In such cases, 
the occupation of $\textit{d}$ orbitals at one site is often correlated with the occupation at the next site. Therefore, analogous to spin ordering, the 
orbital degrees of freedom can be locked and particular orbitals are occupied at each site, which is called orbital ordering. This orbital ordering, 
generally results from the interplay between structural distortion and Coulomb correlation of electrons. Orbital degrees of freedom often couple with 
spin degrees of freedom\cite{1} and give rise to anisotropic electronic and magnetic interactions. Due to strong correlations between different degrees
of freedom, orbital ordering spontaneously drive various exotic phenomena such as reduced dimensionality\cite{1}, structural phase transition\cite{2}, exotic magnetism\cite{3}, Peierls transitions\cite{4}, charge ordering\cite{5} and orbital ordering induced ferroelectricity\cite{6}.

	The possibility of orbital ordering has been discussed for doubly degenerate e$_g$ orbitals of perovskite manganites\cite{8}, cuprates\cite{1,9}
and for the degenerate t$_{2g}$ orbitals of vanadates\cite{10,14}. Most studies investigating this aspect have focused on 3$\textit{d}$ TMOs; much
 less is known about whether related 4$\textit{d}$ and 5$\textit{d}$ oxides can exhibits similar behaviour. Recently, 4$\textit{d}$ and 5$\textit{d}$ 
TMOs have generated significant excitement because of the observation of unexpected orbital ordering\cite{15} and localized transport properties and 
magnetism\cite{16,17} in ruthenate and iridate compounds, respectively. Due to the large spatial extension of 5$\textit{d}$ orbitals, the effects of 
electron correlation are weaker in 5$\textit{d}$ oxides than in 3$\textit{d}$ oxides, while the trend is opposite for the effects of spin-orbit coupling (SOC). Therefore, 5$\textit{d}$ oxides should exhibit a different balance among the effect of spin, orbital, 
and charge degrees of freedom. Although the importance of SOC in magnetic materials has been well known, the interplay between SOC and electronic Coulomb 
interactions in magnetic materials remains a largely unexplored field. The reason for this seems to be the mutually exclusive domain involved in their 
acting. It is only recently that this separation has been questioned\cite{16,17,18,19}. 

\begin{figure}
\begin{center}
\rotatebox{0}{\includegraphics[width=0.52\textwidth]{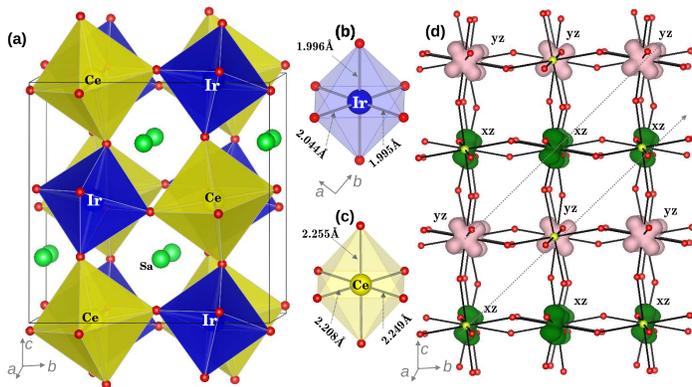}}
\end{center}
\caption{ (Color online)(a) Crystal structure of Sr$_2$CeIrO$_6$. Green, yellow, blue and red spheres represent Sr, Ce, Ir, and O atoms respectively. 
(b) and (c) Bond lengths marked octahedral entities of IrO$_6$ and CeO$_6$ respectively. (d) Three-dimensional magnetization density plot showing the 
orbital ordering. The orbital characters $\textit{d}_{yz}$ (pink surface) and $\textit{d}_{xz}$ (green surface) at each Ir site are marked. The isovalue 
was chosen to be 70 e$^-/\AA^3$. The dashed lines designate the orbital chains in the structure. }
\end{figure}

The 5$\textit{d}$ double perovskite family is one of the most widely investigated oxides for studies exploring exotic phenomena such as magneto-optic 
effect\citep{21}, high-T$_c$ ferrimagnets\cite{22}, half metallicity\cite{23}, peculiar AFM spin structures\cite{24,24b,24c}, as well as strong 
SOC driven phenomena such as spin liquid\cite{25,27}, topological Mott insulator\cite{19,28} and the Weyl semimetal\cite{29}. In the 
context of orbital ordering, 5\textit{d} based systems have been less investigated, because generally for 5\textit{d} elements such as Os, Ir, the 
SOC energy scale considerably exceeds the structural distortion energy scale. Strong SOC destroys orbitally degenerate electronic states required for orbital ordering. In this letter we report a particularly striking example of the interplay among SOC, Coulomb 
interactions and crystal field splitting in the double-perovskite Sr$_2$CeIrO$_6$, which shows a rare type of orbital ordering originating from the 
Ir-t$_{2g}$ states. Using transport and magnetization measurements in addition to density-functional theory based first-principles calculations we 
explored this interesting aspect in the Sr$_2$CeIrO$_6$ double perovskite compound. We propose a rare case of t$_{2g}$ orbitals derived anti-ferro type 
orbital ordering in this material at the Ir site with the half filled (5$\textit{d}^5$) configuration while the Ce remains in the inert [Xe] 
configuration. Our proposed t$_{2g}$ band derived anti-ferro orbital ordering constitute a strong contrast to the most widely known anti-ferro orbital ordering in KCuF$_3$\cite{1} originated from the two e$_g$ orbitals. Moreover only a few frustrated magnetic 
materials have been investigated in the context of  orbital ordering, such as NaNiO$_2$\cite{30}, LiNiO$_2$\cite{31,31b} and spinel 
ZnV$_2$O$_4$\cite{31c}. In this context Sr$_2$CeIrO$_6$ double perovskite is a rare example of a prototype geometrically frustrated fcc lattice where 
orbital ordering plays a crucial role. Our calculations revealed that the experimentally observed AFM insulating ground state is driven by strong SOC 
in addition to small onsite electron-electron Coulomb correlation at the Ir site. These particular behaviours make this system a potential candidate 
for investigating the interplay among the multiple degrees of freedom, such as spin, orbital, lattice, and multiple energy scales e.g. SOC and Coulomb correlations, that interact and compete with each others.

Sr$_2$CeIrO$_6$ crystallized in a monoclinic crystal structure with space group P2$_1$/n, with the lattice parameters a =5.8255 $\AA$, b = 5.8445 
$\AA$, c = 8.2435 $\AA$ and $\beta$= 90.17$^o$, these values are very similar to the reported structure\cite{37}. Due to lack of sensitiveness of  
positions of light atoms such as oxygen in X-ray diffractions based structural analysis, the theoretically optimized structure of Sr$_2$CeIrO$_6$ was obtained by relaxing the atomic positions of all atoms while keeping the lattice parameters fixed at the experimentally determined values. The structure consists of alternating 
corner sharing IrO$_6$ and CeO$_6$ octahedra, in all directions, as shown in Fig. 1(a). 
The Sr atoms are situated at the void positions between the two types of octahedra. Both IrO$_6$ and CeO$_6$ are distorted, the respective six 
metal-oxygens bond lengths are grouped into three values as shown in Figs. 1 (b) and (c) respectively. Interestingly the in-plane 
Ir-O bond lengths of two neighbouring IrO$_6$ octahedra are of opposite type, i.e the short in-plane Ir-O bond of one octahedron corresponds to 
the long in-plane bond of the neighbouring octahedron. Though the positions of the oxygen atoms shifts in the optimized structure compared to the experimental values, the average Ir-O and Ce-O bond lengths of the theoretically optimized structure (2.011 $\AA$ and 2.237 $\AA$ respectively) are in good agreement with the experimental average Ir-O (2.009 $\AA$) and Ce-O (2.202 $\AA$) bond lengths.

The magnetic susceptibilities of Sr$_2$CeIrO$_6$ were measured using a SQUID magnetometer (Quantum Design, MPMS MultiVu Application) in the temperature range from 2 to 300 K in varying field's up to 7 T, see Fig. 2. Inverse magnetic susceptibility data indicated AFM ordering at 
about 21 K. Fitting the high-temperature susceptibility to the Curie-Weiss law [$\chi=\frac{c}{T-\theta}$], gave  $\mu^{eff}$= 1.59 $\mu_B$ with a Weiss 
constant of $\theta$= -108 K, which are very similar to the previously recorded values\cite{37}. A negative value of $\theta$ implies 
dominant AFM interactions in the system. The observed effective magnetic moment indicates, that Ir$^{4+} (\textit{d}^5$) is in S=$\frac{1}{2}$ low-spin 
state and that $\mu^{eff}$ is somewhat smaller than the expected spin-only contribution [$\sqrt{S(S+1)}$] for the S=$\frac{1}{2}$ state. The inset of 
Fig. 2 shows the exponential decay of the resistivity as a function of temperature measured in a 0.01 T field with the Quantum Design PPMS MultiVu 
Application. Fitting the high temperature resistivity data with the semiconducting hopping mechanism, $\rho=\rho_0 e^{\frac{E_g}{2K_BT}}$, resulted in a bulk band gap (E$_g$) of $\approx $ 0.3 eV, indicating small gap insulating behaviour of this material. 

\begin{figure}
\begin{center}
\rotatebox{0}{\includegraphics[width=0.4\textwidth]{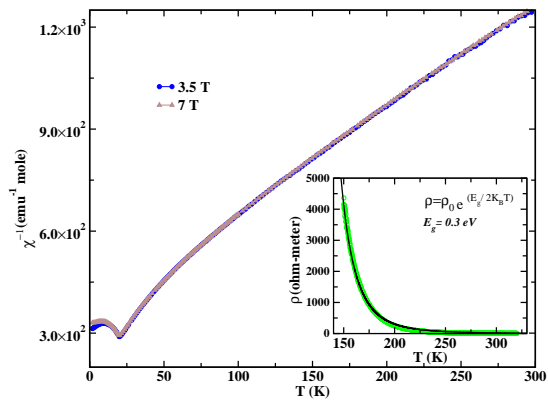}}
\end{center}
\caption{(Color online)Temperature  dependence of the inverse magnetic susceptibility of Sr$_2$CeIrO$_6$ measured at field strengths of 3.5 T and 7 T. 
The inset shows the temperature dependence of the resistivity measured in a 0.01 T field. The green (light) dots represent the experimental data while the solid line represents the fitting with exponential variation.}
\end{figure}

The calculated ferromagnetic (FM), GGA+$\textit{U}$ density of states (DOS) projected on to Ir-5$\textit{d}$ states is shown in the top left panel of 
Fig. 3. The empty Sr states lay far away from the Fermi level (E$_f$), which is consistent with the nominal Sr$^{2+}$ valence state. At the E$_f$ only 
Ir-5$\textit{d}$ states contribute whereas the Ce-4$\textit{f}$ states are empty and situated far above the E$_f$, although Ir-5$\textit{d}$ states 
show significant mixing with the Ce-4$\textit{f}$ and O-2$\textit{p}$ states. We found that in the absence of on-site Coulomb correlation $\textit{U}$, 
the band structure shows a metallic character. The incorporation of small Coulomb correlation $\textit{U}_{eff} \geq$ 2 eV at the Ir site, drove the 
insulating band structure. To clarify the effect of Coulomb correlation $\textit{U}$ on the electronic structure we systematically varied 
$\textit{U}_{eff}^{Ir}$ from 0 to 5 eV, as shown in Fig. S2 (in the supplemental material). The gap, and magnetic moments [0.63 $\mu_B$ ($\textit{U}_{eff}^{Ir}$ 
= 0 eV) to 0.86 $\mu_B$ ($\textit{U}_{eff}^{Ir}$ = 5 eV)] at Ir sites, increased with increasing $\textit{U}$, while Ce stayed in the nonmagnetic state 
with zero magnetic moment. The orbital occupancies and spin states remained unchanged across the entire applied $\textit{U}$ range. Comparing our 
calculated band structure to the resistivity derived band gap, we choose to use $\textit{U}_{eff}^{Ir}$ = 3 eV (which, also is the commonly used $\textit{U}$ value for Ir in the literature\cite{38}), for the following calculations. Because of the octahedral environment the five $\textit{d}$ states were broadly split into t$_{2g}$ ($\textit{d}_{xy}$, $\textit{d}_{yz}$, $\textit{d}_{xz}$) and e$_{g}$ ($\textit{d}_{3z^2-r^2}$, $\textit{d}_{x^2-y^2}$) 
states as marked in the top left panel of Fig. 3. The Ir-t$_{2g}$ states were completely filled in the majority spin channel 
and partially filled in the minority spin channel, whereas the Ir-e$_g$ states were completely empty in both spin channels. These findings in addition to  the calculated magnetic moment suggest that Ce is in the 4+ (Ce$^{4+}$:[Xe]) valence electronic state with zero magnetic moment in the inert configuration, and that Ir is in the 4+ (5$\textit{d}^5$) low spin valence electronic state with S =$\frac{1}{2}$. These findings are consistent with the experimental magnetization measurements and previous electronic structure results\cite{39}. However in the previous studies one failed to drive the insulating band structure without inclusion SOC.

 Because of the  octahedral distortion, the t$_{2g}$ and e$_g$ states are no longer degenerate. Contraction 
of IrO$_6$ octahedra along the $\textit{c}$ axis splits the triply degenerate t$_{2g}$ states into a lower $\textit{d}_{xy}$ band (a$_{1g}$ symmetry), 
and upper $\textit{d}_{yz}-\textit{d}_{xz}$ bands (e$_g^\pi$ symmetry), as shown in the level diagram plot in the right top of Fig. 3. The 
$\textit{d}_{xy}$ band is filled in both spin channels and this is electronically inactive. The two e$_g^\pi$ bands  
are filled by two majority spins and one minority spin, which results S=$\frac{1}{2}$ state. However the single minority spin has the freedom to occupy 
any orbital between the double degenerate $\textit{d}_{yz}-\textit{d}_{xz}$ bands. Therefore this minority spin experiences orbital degeneracy and 
drive a definite orbital ordering pattern in the lattice. To clarify this orbital degeneracy we plotted the orbital projected DOS for two Ir sites, Ir-1 
(0.5, 0.0, 0.5) and Ir-2 (0.0, 0.5, 0.0), in the unit cell, as shown in the left middle and lower panels of Fig. 3 respectively. The plots show that the 
$\textit{d}_{yz}$ state of the Ir-1 site is equivalent to the $\textit{d}_{xz}$ state of the Ir-2 site, i.e.  at the Ir-1 site the $\textit{d}_{xz}$ 
state is occupied while for the Ir-2 site, the $\textit{d}_{yz}$ state is populated by the single minority spin in the e$_g^\pi$ bands, which is clearly 
shown in the energy level diagram of the respective right panels. To visualize the pattern of orbital ordering we plotted the magnetization density within GGA+$\textit{U}(\textit{U}_{eff}^{Ir}$ = 3 eV) as shown in Fig.1(d). This figure shows that crystallographically two non-equivalent Ir sites, i.e 
Ir-1 and Ir-2 in the unit cell are decorated with the alternating $\textit{d}_{yz}$ and $\textit{d}_{xz}$ type orbitals and that these alternating orbital chains run along the body diagonal to the monoclinic unit cell. Because in this case the Ir sites prefer two different types of orbitals, we described this type of orbital ordering as anti-ferro type orbital ordering. Interesting point to be noted is that, aforementioned anti-ferro orbital ordering consisting of two t$_{2g}$ orbitals is in contrast to the most widely known classic example of anti-ferro orbital ordering as found in KCuF$_3$\cite{1}, where alternating sites of Cu are occupied by the two e$_g$ orbitals. Therefore Sr$_2$CeIrO$_6$ is a very unique case.

The resulted orbital ordering could be understood as an effect of the cooperative Jahn-Teller distortion associated to the IrO$_6$ octahedra. The individual IrO$_6$ octahedra are distorted as reflected by the three different types of Ir-O bond lengths in each octaherdon. If we concentrate only on 
the $\textit{ab}$ plane Ir-O bond lengths, theses are grouped into two Ir-O bond lengths as short 1.986 $\AA$ and long 2.033 $\AA$. However the 
crystallographic orientation of short and long bond lengths for the two Ir sites (i.e Ir-1 and Ir-2) are opposite. For Ir-1 the short bond (1.986 $\AA$) is along the crystallographic $\textit{b}$ axis and the long bond (2.033 $\AA$) along the crystallographic $\textit{a}$ axis, while for Ir-2, short bond is along the crystallographic $\textit{a}$ axis and long bond along the crystallographic $\textit{b}$ axis, i.e short bond of the Ir-1 octahedron
corresponds to the long bond of the Ir-2 octahedron, see right middle and bottom panel of Fig. 3. Therefore this cooperative Jahn-Teller distortion in the IrO$_6$ octahedra, in terms of crystallographic orientation of Ir-O bond lengths, induces the occupation of two different orbitals in the two neighbouring distorted Ir octahedra. 

\begin{figure}
\begin{center}
\rotatebox{0}{\includegraphics[width=0.55\textwidth]{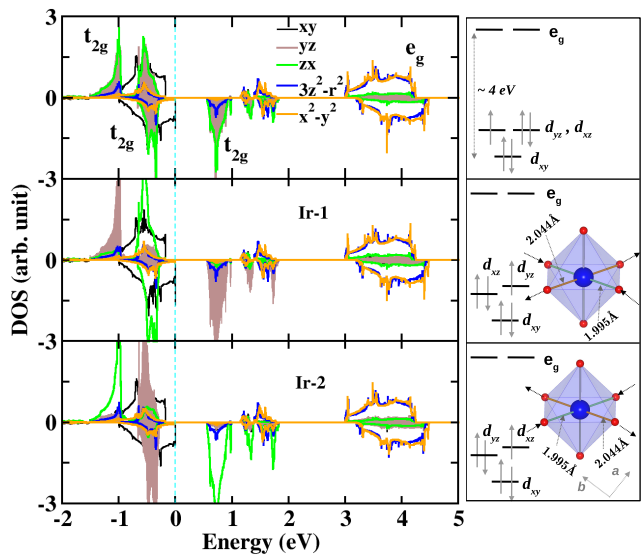}}
\end{center}
\caption{(Color online) GGA+$\textit{U}(\textit{U}_{eff}^{Ir}$ = 3 eV) FM spin-polarized Ir-5$\textit{d}$ orbital projected DOS. The two channels of the all panels  represent majority and minority spin channels. The top panel represents the total Ir-$\textit{d}$ DOS in the unit cell while middle and lower 
panels represent the orbital projected DOS separately for the two Ir sites in the unit cell. The Fermi level is marked at zero on the energy scale. The 
right panels represent the energy level diagram corresponding to the left panel DOS. The individual IrO$_6$ octahera in the right panels are shown to 
clarify the bond length orientation in the crystallographic axes. In plane short (1.986 $\AA$) and long (2.033 $\AA$) bonds were marked by green and 
orange respectively. The arrow indicates the contraction and elongation of bond length along crystallographic axes.}
\end{figure}

To understand the role of SOC in this orbitally active material, we performed GGA+$\textit{U}$+SOC ($\textit{U}_{eff}^{Ir}$ = 1-5 eV) calculations. 
Surprisingly we found that GGA+SOC without onsite $\textit{U}$, could not drive the insulating nature of the band structure. We indeed required a small 
onsite correlation ($\textit{U}_{eff}^{Ir} \geq$ 1 eV) along with SOC to open an insulating gap in the band structure. Because of the orbital 
degeneracy 
of Ir, we found that orbital magnetic moment to be larger than the spin magnetic moment at the Ir site, although the value of the orbital and spin 
magnetic moments depended sensitively on chosen $\textit{U}$ values. For example, as we increased from a small value of $\textit{U}_{eff}^{Ir}$ = 1 eV 
[$\mu^{spin} (\mu^{orbital}$)=0.25 (0.32) $\mu_B$/Ir] to moderate value of $\textit{U}_{eff}^{Ir}$ = 3 eV [$\mu^{spin} (\mu^{orbital}$) = 0.33 (0.43) 
$\mu_B$/Ir] to a large value of $\textit{U}_{eff}^{Ir}$ = 5 eV [$\mu^{spin} (\mu^{orbital}$)=0.38 (0.53) $\mu_B$/Ir], the spin, orbital moment and the 
ratio of $\mu^{orbital}/\mu^{spin}$ [1.28 (for $\textit{U}_{eff}^{Ir}$= 1 eV), 1.34 (for $\textit{U}_{eff}^{Ir}$= 3 eV), 1.39 (for 
$\textit{U}_{eff}^{Ir}$ =5 eV)] increased. We note that the orbital magnetic moments point in the same direction as the spin magnetic moment which 
is expected for the more than half filled t$_{2g}$ occupancies at the Ir sites. Similarly large orbital magnetic moment and a similar 
$\mu^{orbital}/\mu^{spin}$ ratio are reported in the literature in the context of an other Ir based perovskite BaIrO$_3$\cite{40}. 
We found that, in the GGA+$\textit{U}$ ($\textit{U}_{eff}^{Ir}$=0-5 eV) calculations, FM alignment of Ir spins was energetically lower than the 
different possible AFM configurations. However experiments indicate a clear AFM transition at 17 K. Following the observation of significant orbital 
moments at both Ir sites, we investigated the possible role of SOC in the magnetic ground states. For this purpose, we conducted GGA+$\textit{U}$+SOC 
calculations for FM and different AFM spin configurations. We found that, with the inclusion of SOC, compared to the other possible FM and AFM 
configurations, the A-type AFM state became the energetically lowest [$\mid E_{A-AFM} - E_{FM} \mid$= 2.8 meV/f.u] magnetic state. Our findings  perfectly 
agree with previous electronic structure studies \cite{39} as they also mentioned SOC driven stabilisation of the AFM state. In this A-AFM configuration, 
Ir spins are aligned ferromagnetically in the $\textit{ab}$ plane, and were coupled antiferromagnetically in the out-of-plane direction. We note that 
the experimental band gap ($\approx $ 0.3 eV) agrees qualitatively with the GGA+$\textit{U}$+SOC ($\textit{U}_{eff}^{Ir}$ = 3 eV) AFM band gap 
($\approx$ 0.35 eV). In Fig. S3 (in the supplementary material), we show the GGA+$\textit{U}$ and GGA+$\textit{U}$+SOC band dispersion for the FM and AFM 
configurations. Because of the increase of Fermi energy for the AFM state over the FM state with the inclusion of SOC, the experimentally observed AFM 
state becomes energetically lower than the FM state. We confirmed the stability of the AFM state over FM configurations by cross checking our findings 
across the range of $\textit{U}_{eff}^{Ir}$ values from 1 to 5 eV, and we verified the existence of the proposed orbital ordering in the presence of AFM 
coupling as well. An interesting point to note is that, to obtain the correct magnetic ground state, the inclusion of SOC was mandatory, which reflects 
the importance of SOC in this Ir compound. We note a similar mechanism was recently suggested for other iridates\cite{39,41}. Therefore this material may 
be classified as a  Coulomb enhanced SOC driven AFM insulator.

It is common believed that strong SOC destroys orbital ordering, however, the present material shows a rare exception. To understand the possible role of 
competition between SOC and structural distortion in deriving orbital ordering, we compare our results with the isoelectronic  Ba variant 
(Ba$_2$CeIrO$_6$) of the present compound. Since Ba$^{2+}$ has a larger radius the for the  tolerance factor for the later (t=0.991) is close to 1. Therefore Ba form in an ideal cubic crystal structure \cite{42} and the electronic structure does not allow a$_{1g}$-e$_g^{\pi}$ splitting required 
for orbital ordering. Because of the smaller radius of Sr$^{2+}$, the tolerance factor (t=0.935) of Sr$_2$CeIrO$_6$ significantly deviates from 1 and thus a distorted monoclinic crystal structure results that allows the orbital ordering. From the total energy calculations we found that for Ba$_2$CeIrO$_6$, without inclusion of SOC, the monoclinic structure is lower (by 138 meV/f.u) than the cubic and we found similar orbital ordering as Sr$_2$CeIrO$_6$. However, inclusion of  SOC cubic structure became $\approx$ 17 meV/f.u lower than the monoclinic. In contrast for the Sr$_2$CeIrO$_6$ both without and with SOC, the monoclinic structure is energetically $\approx$ 747 meV/f.u lower than than the cubic one. These analyses gave 
indirect evidence that, since SOC energy scale remains unchanged between two compounds, in Sr$_2$CeIrO$_6$ the structural distortion is strong 
enough to overcome the negative SOC effect and we do obtain the orbital ordering state, which appears not to be possible for Ba$_2$CeIrO$_6$.

In conclusion, we conducted both experiments and DFT based first-principles calculations to investigate the nature of the orbital ordering in the ordered 
double perovskite Sr$_2$CeIrO$_6$, with Ir as a single electronically active site in the geometrically frustrated fcc lattice. We found t$_{2g}$ band 
derived anti-ferro type of orbital ordering at the Ir site consisting of $\textit{d}_{yz}$ and $\textit{d}_{xz}$ orbitals that were associated by 
to alternating Ir sites along the body diagonal to the unit cell. Further experiments, such as X-ray resonant spectroscopy, would be helpful to directly 
probe our proposed orbital ordering pattern. We found that small electron-electron onsite correlation $\textit{U}$ is also relevant in this 
5$\textit{d}$-Ir system for obtaining the correct insulating electronic structure. SOC played a very crucial role in this orbitally active material, for 
stabilizing the experimentally observed AFM ground state in addition to the small correlation effect, and with a large contribution of orbital magnetic 
moment greater than the spin counterpart. This Ir based double perovskite material has a very striking feature in which both electron-electron 
interaction and SOC energy scales are important for understanding the electronic and magnetic behaviour. More detailed experiments and theoretical 
analysis are  required to explore this critical energy balance in this interesting material.

We thank the helpful discussions with Prof. J. van der Brink. This work is financially supported by the ERC Advanced Grant (Grant No. 291472)\

$^*$ m.jansen@fkf.mpg.de \


\end{document}